\documentclass[aps,prd,twocolumn,showpacs,showkeys,amsmath,amssymb]{revtex4}
\usepackage{graphicx}
\usepackage{dcolumn}
\usepackage{bm}
\begin{document}
\title{Spin ${3\over 2}$ Pentaquarks}

\author{W. Wei, P.-Z. Huang, H.-X. Chen}
\affiliation{%
Department of Physics, Peking University, BEIJING 100871, CHINA}
\author{Shi-Lin Zhu}
\email{zhusl@th.phy.pku.edu.cn} \affiliation{ RCNP, Osaka
University, Japan}
\affiliation{ Department of Physics, Peking
University, BEIJING 100871, CHINA}

\begin{abstract}

We investigate the possible existence of the spin 3/2 pentaquark
states using interpolating currents with K-N color-octet structure
in the framework of QCD finite energy sum rule (FESR). We pay
special attention to the convergence of the operator product
expansion.

\end{abstract}


 \maketitle

\pagenumbering{arabic}

\section{Introduction}\label{sec1}

Since LEPS collaboration reported the $\Theta^+$ pentaquark
\cite{lep}, many subsequent experiments claimed the confirmation
of this exotic baryon with S=+1 while many other groups didn't
observe it in their search. Experiments with both positive and
negative results are reviewed in Refs. \cite{hicks,ddd}. At
present, neither the existence nor the non-existence of $\Theta^+$
is established, which can only be settled by the new
high-statistical experiments.

If $\Theta^+$ {\sl really} exists, its low-lying mass, extremely
narrow width and weird production mechanism pose a serious
challenge to theorists. There have been over four hundred
theoretical papers addressing these issues. The early theoretical
development can be found in a recent review \cite{slz2}. The
angular moment of $\Theta^+$ is assumed to be $J={1\over 2}$ in
order to render $\Theta^+$ low-lying in most of these efforts.
Moreover, some models prefer the positive parity to ensure a
narrow $\Theta^+$.

However, the possibility of $J=\frac 32$ is not excluded. For
example, $J^P=\frac 12 ^+$ pentaquarks are always accompanied by
$J^P=\frac 32 ^+$ partners \cite{dudek} in Jaffe and Wilcezk's
diquark model \cite{jw1}. Their magnetic moments were calculated
in \cite{liww} while their radiative decays and photoproduction
were studied in \cite{he}.

If $\Theta^+$ carries $J^P={3\over 2}^-$, it decays into nucleons
and kaons via D-wave. The phase space suppression factor is $\sim
\left( {p_K\over m_\Theta}\right)^5 \sim 10^{-5}$ where $p_K$ is
the decay momentum of the kaon. The decay width could be well
below ten MeV even if the coupling constant $G_{\Theta K N}$ is
big due to the absence of the orbital excitation inside $\Theta^+$
and $\Theta$'s strong overlap with $K N$.

There have been a few theoretical papers on the possible
$J={3\over 2}$ pentaquarks using different models. Page and Robert
suggested $I=2, J^P={1\over 2}^-, {3\over 2}^-$ for $\Theta^+$ to
resolve the narrow width puzzle \cite{page}. Jaffe and Wilczek
discussed the $J^P={3\over 2}^-$ assignment for $\Xi$ pentaquark
\cite{jaffe2}. The mass spectrum of $J^P={3\over 2}^-$ pentaquarks
were studied with the perturbative chiral quark model
\cite{faessler}. Takeuchi and Shimizu suggested the observed
$\Theta$ resonance as a $I=0, J^P=\frac 32 ^-$ $N K^\ast$ bound
state using the quark model \cite{takeuchi}. With the flux tube
model, Kanada-En¡¯yo et al. studied the mass and decay width of
the $I=1, J^P=\frac 32 ^-$ pentaquark \cite{kanada}. Huang et al.
proposed $\Theta ^+$ as a molecular state of $NK{\pi}$ with $I=1,
J^P=\frac 32 ^- $ using the chiral SU(3) quark model
\cite{zyzhang}. The phenomenology of $J={3\over 2}$ pentaquarks
such as the mixing scheme and mass pattern was discussed in
\cite{hyodo}. Pentaquark states with J=3/2 and I=0,1 were studied
using currents composed of one scalar diquark and one vector
diquark \cite{qsr-nishi}.

We shall employ QCD sum rules (QSR) to explore the possible
existence of the $J={3\over 2}, S=+1$ pentaquark states with the
isospin $I = 0, 1, 2$. QSR formalism was first employed to study
the pentaquark mass with different isospin in Ref. \cite{qsr-zhu}.
Up to now there have been more than ten papers on pentaquarks
within this framework
\cite{qsr-mat1,qsr-oka,hpz,qsr-eide,qsr-mat2,qsr-nava,qsr-oga,
qsr-mori1,qsr-su,qsr-nishi,narison,qsr-lee,wsl}.

In practice there are two forms of QCD sum rules. The first one is
the conventional Laplace sum rules introduced originally by the
inventors of this formalism \cite{svz}. The other one is the
finite energy sum rule (FESR) \cite{fesr}. Their difference lies
in the weight function. The right-hand-side (RHS) of the
traditional sum rules deals with $ \int^{s_0}_{s_{min}} \rho (s)
e^{-{s/ M_B^2}} ds $, where $\rho(s)$ is the spectral density
including the nonperturbative power corrections arising from
various condensates. $M_B$ is the Borel parameter. $s_{min}$ is
the starting point of the integral, which is zero for massless
quarks. In the analysis of the sum rules, the quark-hadron duality
is always invoked. Starting from $s_0$, the physical spectral
density, which arises from the higher resonances and continuum is
always replaced by the perturbative one. Hence $s_0$ is called the
threshold parameter and is typically around the radial excitation
mass.

For the FESR approach, the exponential weight function
$e^{-{s/M_B^2}}$ is replaced by $s^n$ in the numerical analysis.
For the conventional ground-state hadrons such as the rho and
nucleon, both the Laplace QSR and FESR yield almost the same
numerical results for the hadron mass, thanks to (1) the good
convergence of the operator product expansion; and (2) the useful
experimental guidance on the threshold parameter $s_0$. The reason
is simple: the rough value of the radial excitation is more or
less known experimentally.

In the present case, the situation is more tricky. Even the
existence of the lowest $\Theta^+$ pentaquark has not been
established, let alone its radial excitation. On the other hand,
the spectral density $\rho(s) \sim s^m$ with $m\ge 5$, which
causes strong dependence on the continuum or the threshold
parameter $s_0$. Hence, compared to the Laplace sum rule with the
double parameters $(s_0, M_B)$, the single-parameter FESR may have
some advantage as recently noted in \cite{narison}. With FESR, one
can study the dependence of $m_\Theta$ on $s_0$ in the working
region.

On the other hand, the disadvantage of FESR is that its weight
function $s^n$ enhances the continuum part even more than the
weight function $e^{-s/M_B^2}$ in the Laplace sum rule. Hence some
uncertainty is connected to the continuum threshold $s_0$.
Especially one must make sure that only the lowest pole
contributes to the FESR below $s_0$. Otherwise the result will be
misleading, which will be shown explicitly in our numerical
analysis. To be more specific, a {\sl naive} stability region in
$s_0$ is no guarantee of a {\sl physically reasonable} value for
$s_0$. For example, the FESR with an extracted threshold
$s_0\approx 20$ GeV$^2$ is certainly irrelevant for the $\Theta^+$
pentaquark around 1.53 GeV.

If a narrow resonance {\sl really} exists, there should exist (1)
a spectral density $\rho (s)$ with reasonably good behavior and
(2) some values of $s_0$, on which the dependence of $m_\Theta$ is
weak. For example, an oscillating $\rho (s)$ from negative to
positive values around $m_\Theta^2$ is regarded as having "bad"
behavior. The physical spectral density should take the
Breit-Wigner form around $m_\Theta^2$ if $\Theta^+$ {\sl really}
exists as a very narrow resonance. Hence, $\rho (s)$ should be
either fully positive-definite or negative-definite around
$m_\Theta^2$.

We will use FESR to analyze the possible existence of $J={3\over
2}, S=+1$ pentaquarks in this work. This paper is organized as
follows. In Section \ref{sec2}, we construct the interpolating
currents with different isospin and present the formalism. The
spectral densities and numerical analysis are given in Section
\ref{sec3}. The last section is a short discussion.

\section{Formalism of FESR}\label{sec2}

The method of QCD sum rules \cite{svz,ioffe,reinders} incorporates
two basic properties of QCD in the low energy domain: confinement
and approximate chiral symmetry and its spontaneous breaking. One
considers a correlation function of some specific interpolating
currents with the proper quantum numbers and calculates the
correlator perturbatively starting from high energy region. Then
the resonance region is approached where non-perturbative
corrections in terms of various condensates gradually become
important. Using the operator product expansion, the spectral
density of the correlator at the quark gluon level can be obtained
in QCD. On the other hand, the spectral density can be expressed
in term of physical observables like masses, decay constants,
coupling constants etc at the hadron level. With the assumption of
quark hadron duality these two spectral densities can be related
to each other. In this way one can extract hadron masses etc.

We use the following interpolating current for the $I =0, S=+1,
J={3\over 2}$ pentaquark state
\begin{eqnarray}\label{current1}\nonumber
 \eta^0_{\mu,K-N}(x)=\frac{1}{\sqrt{2}}\varepsilon^{abc}
\Big[u^T_a(x)C\gamma_5d_b(x)\Big]\Big\{u_e(x)\bar{s}_e(x)\times &\\
\gamma_\mu d_c(x)
 -d_e(x)\bar{s}_e(x)\gamma_\mu u_c(x)\Big\}
\end{eqnarray}
where three quarks and the remaining $\bar q q$ pair are both in a
color adjoint representation \cite{qsr-zhu}. Similarly, we can
introduce the $I =1, J={3\over 2}$ current
\begin{eqnarray}\label{current2}\nonumber
 \eta^1_{\mu,K-N}(x)=\frac{1}{\sqrt{2}}\varepsilon^{abc}
\Big[u^T_a(x)C\gamma_5d_b(x)\Big]\Big\{u_e(x)\bar{s}_e(x)& \\
\gamma_\mu d_c(x)+d_e(x)\bar{s}_e(x)\gamma_\mu u_c(x)\Big\} \;.
\end{eqnarray}
For $I=2, I_z=2, J={3\over 2}$ state, we use
\begin{equation}\label{current4}
\eta^2_{\mu,K-N}(x)=
\Big[u^T_a(x)C\gamma_{\nu}u_b(x)\Big]\gamma^{\nu}\gamma_5u_e(x)\bar{s}_e(x)\gamma_\mu
u_c(x)\;.
\end{equation}

The overlapping amplitude $f_j$ of the interpolating current is
defined as
\begin{equation}
\langle 0| \eta_{\mu} (0)| \frac 32,p,j \rangle = f_j
\upsilon_{\mu}(p)
\end{equation}
where $j$ is the isospin. $\upsilon_{\mu}$ is the Rarita-Schwinger
spinor for the $J={3\over 2}$ pentaquark, which satisfies
$(\hat{p}-M_X)\upsilon_{\mu}=0$,
${\bar\upsilon_{\mu}}\upsilon^{\mu}=-2M_X$, and
$\gamma_{\mu}\upsilon^{\mu}=p_{\mu}\upsilon^{\mu}=0$.

We consider the following correlation function
\begin{eqnarray}\label{cor}\nonumber
i\int d^4xe^{-ipx}<0|T\{\eta_{\mu}(x) \bar{\eta_{\nu}}(0)\}|0>=
&\\
g_{\mu \nu}\left(\Pi_{A}(p^2)\hat{p}+\Pi_{B}(p^2) \right)+ \cdots
\end{eqnarray}
where the ellipse denotes other Lorentz structures which receive
contributions from both $J={1\over 2}$ and $J={3\over 2}$
resonances. The tensor structures $g_{\mu \nu}, g_{\mu
\nu}\hat{p}$ are particular. They receive contribution only from
the $J={3\over 2}$ pentaquarks.

We can write a dispersion relation for the scalar functions
$\Pi^{A, B}(p^2)$.
\begin{equation}\label{disp}
\Pi_{A, B}(p^2)=\int ds {\rho_{A,B}(s) \over (s-p^2 -i \epsilon )
}
\end{equation}
where $\rho_{A,B} (s)$ is the spectral density.

The width of the $\Theta^+$ pentaquark is less than several MeV.
For such a narrow resonance, its spectral density can be
approximated by a delta function very well. Hence, at the hadron
level we have
\begin{eqnarray}
\rho_{A}(s)= f_A^{2} \delta (s-M_j^2) + \mbox{higher states}
\\ \nonumber
\rho_{B}(s)= f_B^{2} M_j \delta (s-M_j^2) + \mbox{higher states}
\end{eqnarray}
where $M_j$ is the pentaquark mass. In principle, there also
exists non-resonant kaon nucleon continuum contribution to
$\rho_{A, B}(s)$ since $\Theta^+$ lies above threshold. However,
this kind of non-resonant K N continuum is either of D-wave for
$J^P={3\over 2}^-$ or of P-wave for $J^P={3\over 2}^+$. Their
contribution is strongly suppressed compared to the resonant
$\Theta^+$ pole contribution. Here we want to emphasize that
\begin{equation}
f_B^2 = + f_A^{2}
\end{equation}
for $J^P={3\over 2}^+$ pentaquarks while
\begin{equation}
f_B^2 = - f_A^{2}
\end{equation}
for $J^P={3\over 2}^-$ pentaquarks. Hence the relative sign
between $f_B^2$ and $|f_A|^{2}$ indicates the parity of the
corresponding pentaquark.

On the other hand, the spectral density $\rho_{A, B}(s)$ can be
calculated at the quark gluon level. For example, the correlation
function for the interpolating current (\ref{current1}) reads
\begin{eqnarray}\nonumber
&&<0|T\{\eta^1_{\mu}(x) \bar{\eta}^{1}_{\nu}(0)\}|0> =\\ \nonumber
&&\varepsilon^{abc}\varepsilon^{a'b'c'}
\{2iS_u^{ea'}C\gamma_5S_d^{Tcb'}\gamma^T_\mu
S_s^{Te'e}(-x)\gamma^T_\nu S_d^{Tbc'}C\gamma_5S_u^{ae'}\\
\nonumber &&
-iS_u^{ea'}C\gamma_5S_d^{Tbb'}C\gamma_5S_u^{ac'}\gamma_\nu
S_s^{e'e}(-x)\gamma_\mu S_d^{ce'} \\
\nonumber &&-iS_u^{ec'}\gamma_\nu S_s^{e'e}(-x)\gamma_\mu
S_d^{cb'}C\gamma_5S_u^{Taa'}C\gamma_5S_d^{be'}\\  \nonumber &&
+i{\bf\mbox{Tr}}\Big[S_u^{Taa'}C\gamma_5S_d^{bb'}C\gamma_5\Big]S_u^{ec'}\gamma_\nu
S_s^{e'e}(-x)\gamma_\mu S_d^{ce'}\\  \nonumber
&&-iS_u^{ea'}C\gamma_5S_d^{Tbb'}C\gamma_5S_u^{ae'}{\bf
\mbox{Tr}}\Big[S_s^{e'e}(-x)\gamma_\mu S_d^{cc'}\gamma_\nu\Big]\\
\nonumber
&&+i{\bf\mbox{Tr}}\Big[S_u^{Taa'}C\gamma_5S_d^{bb'}C\gamma_5\Big]{\bf
\mbox{Tr}}\Big[S_s^{e'e}(-x)\gamma_\mu
S_d^{cc'}\gamma_\nu\Big]S_u^{ee'}\\  \nonumber &&-i{\bf
\mbox{Tr}}\Big[S_u^{Taa'}C\gamma_5S_d^{bc'}\gamma_\nu
S_s^{e'e}(-x)\gamma_\mu S_d^{cb'}C\gamma_5\Big]S_u^{ee'}\}
\end{eqnarray}
where $S(x)=-i\langle 0|T\{q(x) \bar q(0)\}|0\rangle $ is the full
quark propagator in the coordinate space. Throughout our
calculation, we assume the up and down quarks are massless. The
first few terms of the quark propagator is
\begin{eqnarray}\nonumber
iS^{ab}(x)=\frac{i\delta^ab}{2\pi^2x^4}\hat{x}+\frac{i}{32\pi^2}
\frac{\lambda^n_{ab}}{2}g_sG^n_{\mu\nu}
\frac{1}{x^2}(\sigma^{\mu\nu}\hat{x}+\hat{x}\sigma^{\mu\nu})& \\
\nonumber -\frac{\delta^{ab}}{12}\langle\bar
qq\rangle+\frac{\delta^{ab}x^2}{192}\langle g_s\bar q\sigma
Gq\rangle+\cdots
\end{eqnarray}

First the correlator is calculated in the coordinate space. Then
$\Pi_{A, B}(p^2)$ can be derived after making Fourier
transformation to $\Pi_{A, B}(x)$. From the imaginary part of
$\Pi_{A, B}(p^2)$ one can extract the spectral density $\rho_{A,
B}(s)$ at the quark hadron level.


With the spectral density, the $n$th moment of FESR is defined as
\begin{equation}
W_{A,B}(n,s_0)=\int_{m_s^2}^{s_0}ds s^n \rho_{A, B}(s)
\end{equation}
where $n\ge 0$. With the quark hadron duality assumption we get
the finite energy sum rule
\begin{equation}\label{12}
W_{A,B}(n,s_0)|_{Hadron} =W_{A,B}(n,s_0)|_{QCD}\; .
\end{equation}
The mass and $f_j^2$ can be obtained as
\begin{equation}
M_j^2={W_{A,B}(n+1,s_0)\over W_{A,B}(n,s_0)}
\end{equation}
\begin{equation}
f_j^2 M_j=W_B(n=0,s_0)|_{QCD}\; .
\end{equation}
In principle, one can extract the threshold from the requirement
\begin{equation}
{d M_j^2\over d s_0} =0\; ,
\end{equation}
or
\begin{equation}
\int_{m_s^2}^{s_0}(s_0-s) s^n \rho (s) =0\; .
\end{equation}
If $\rho(s)>0$ or $\rho(s) <0$ in the whole region $[m_s^2,
\infty)$, there does not exist a stable threshold for this finite
energy sum rule .

For the gluon condensates we keep only D=4 term and neglect D=6, 8
pieces in our calculation. The contribution of D=6, 8 gluon
condensates was found to be much smaller than D=4 term in previous
QSR analysis. For D=7-9 power corrections, we keep only those
numerically large terms, which are related to the quark condensate
$\langle\bar qq \rangle$ or the quark gluon mixed condensate
$\langle g_s\bar q\sigma G q\rangle$. Condensates such as
$\langle\bar qq \rangle \langle g_s^2GG\rangle$, $\langle g_s\bar
q\sigma G q\rangle \langle g_s^2GG\rangle$ are neglected.

We use the following values of condensates in the numerical
analysis: $\langle\bar qq \rangle=-(0.24 \mbox{GeV})^3,
\langle\bar ss\rangle=-(0.8\pm 0.1)(0.24 \mbox{GeV})^3, \langle
g_s^2GG\rangle =(0.48\pm 0.14) \mbox{GeV}^4, \langle g_s\bar
q\sigma G q\rangle=-m_0^2\times\langle\bar qq\rangle $,
 $m_0^2=(0.8\pm0.2)$GeV$^2$. We use $m_s(1\mbox{GeV})=0.15
\mbox{GeV}$ for the strange quark mass in the $\overline{MS}$
scheme.

\section{Numerical Analysis}\label{sec3}

\subsection{$I=0$ FESR from $g_{\mu\nu}$ Structure}

After tedious calculation, the spectral density $\rho^0_{B}(s)$
with condensates up to dimension 9 reads
\begin{eqnarray}\nonumber
\rho^0_{B}(s)&=&{1\over 2^{19}\cdot 175\cdot \pi^8}s^5 m_s\\
\nonumber && -{5\over 2^{16}\cdot 27\cdot \pi^6}s^4({\langle\bar
qq\rangle}+\frac{3}{20}{\langle\bar ss \rangle})\\
\nonumber &&+{7\over 2^{20}\cdot 45\cdot\pi^8}s^3m_s{\langle
g_s^2GG\rangle}\\
\nonumber &&  -{17\over 2^{14}\cdot 45\cdot \pi^6}s^3
\langle\bar q\sigma\cdot Gq\rangle \\
\nonumber && +{1\over 2^{12}\cdot 3 \cdot \pi^4}s^2m_s( \frac
{103}{6}{\langle\bar qq\rangle}-5{\langle\bar
ss\rangle}){\langle\bar qq\rangle}\\  &&
 -{1\over 2^{8}\cdot 27\cdot \pi^2}s({56\langle\bar
qq\rangle}+75{\langle\bar ss \rangle}){\langle\bar qq\rangle}^2
\end{eqnarray}
where we have used the factorization approximation for the
high-dimension quark condensates.

Both the perturbative piece and the D=3 power correction from the
quark condensate in $\rho_B^0(s)$ are positive. There are two
types of D=5 power corrections. One arises from the gluon
condensate $m_s\langle g_s^2GG\rangle$. Its contribution is
positive. The other one is from the quark gluon mixed condensate
$\langle g_s\bar q\sigma G q\rangle$, which overwhelms the D=3
quark condensate and D=4 gluon condensate in magnitude and carries
a minus sign. As can be seen from Fig. 1, the quark gluon mixed
condensate renders the spectral density negative for a big range
of s. The D=7 condensate yields a positive contribution and
cancels the big negative contribution from the quark gluon mixed
condensate, leading to a nearly vanishing $\rho_B^0 (s)$ for $s\le
4$ GeV$^2$ to this order. The contribution from the D=9 condensate
is also large. In fact, $\rho_B^0(s)>0$ is positive throughout the
whole range $[m_s^2, \infty)$ with the inclusion of the D=9 power
correction.

\begin{figure}[hbt]\label{bi0rho}
\begin{center}
\scalebox{0.8}{\includegraphics{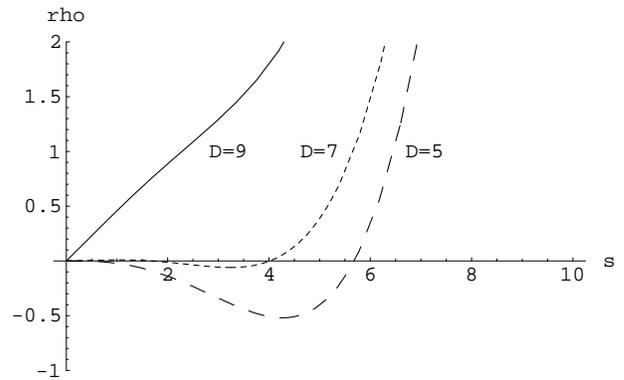}}
\end{center}
\caption{The variation of the spectral density $\rho^0_B$ from the
tensor structure $g_{\mu\nu} $ (in unit of $\mbox{GeV}^{11}\times
10^{-8}$) with s (in unit of $\mbox{GeV}^2$) for the I=0 current
(\ref{current1}). The long-dashed, short-dashed and solid curves
correspond to $\rho^0_B$ with $D=5, 7, 9$ condensates
respectively. }
\end{figure}

It is a common feature in the pentaquark sum rules that the quark
gluon mixed condensate plays a very striking, sometimes dominant
role when the sum rule is truncated at low orders. In contrast,
the ratio between the power corrections from $\langle g_s\bar
q\sigma G q\rangle$ and $\langle\bar qq \rangle$ is less than
$-20\%$ in the nucleon mass sum rule from ${\hat p}$ structure
\cite{ykc}, which ensures the convergence of the operator product
expansion (OPE).

In order to analyze the OPE convergence in the present case, we
list the right hand side (RHS) of Eq. (\ref{12}) for the case of
$n=0$ below:
\begin{equation}\label{ww}
W^{I=0}_B(0,s_0)\sim 3.2\times 10^{-3} s_0+1-{7.3\over
s_0}+{12\over s_0^2}+{247\over s_0^3}
\end{equation}
where the individual terms are normalized according to the quark
condensate.

In the FESR framework the expansion parameter is $1/s_0$ while
it's $1/M_B^2$ in the conventional Laplace sum rule analysis. Eq.
(\ref{ww}) indicates that the present FESR is very sensitive to
the high dimension condensates with $D\ge 5$. Numerically, these
nonperturbative power corrections converge for $s_0\ge 10$GeV$^2$
only. Such a big threshold is irrelevant for the $\Theta^+$
pentaquark around 1.53 GeV. We have to conclude that this FESR is
not suitable for the extraction of pentaquark mass.

If we ignore the convergence problem, the variation of the
pentaquark mass with the continuum threshold is shown in Fig 2.
Naively, there exists a "stable" threshold of $s_0=5.6$ GeV$^2$
corresponding to $m_\Theta=(1.8\pm 0.1)$ GeV if we truncate the
sum rule at D=5 order.

\begin{figure}[hbt]
\begin{center}
\scalebox{0.8}{\includegraphics{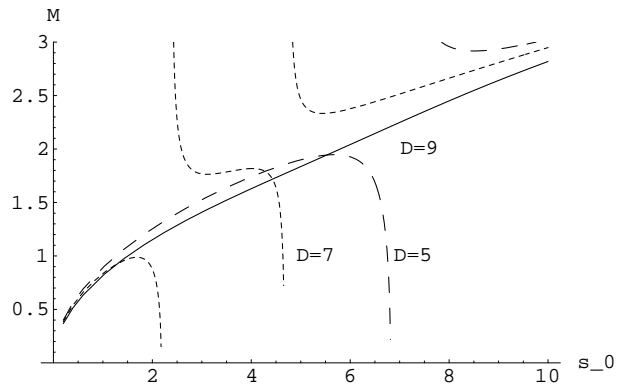}}
\end{center}\label{bi0m}
\caption{The variation of the mass $M$ with the threshold $s_0$
(in unit of $\mbox{GeV}^2$) corresponding to $\Pi_B$ of the I=0
current (\ref{current1}). Same conventions as in Fig. 1.}
\end{figure}

However, if we add the D=7 correction, the shape of the curve
changes dramatically. With the inclusion of D=9 condensates, there
is no stable continuum threshold at all. The underlying reason of
the extreme sensitivity of this FESR to the high dimension
condensates is the lack of the convergence of the operator product
expansion. Hence the extracted pentaquark mass and continuum
threshold are very unreliable.

\subsection{$I=0$ FESR from $g_{\mu\nu}{\hat p}$ Structure}

Now let's move to the chirally even tensor structure
$g_{\mu\nu}{\hat p}$. $\rho^0_{A}(s)$ with condensates up to
dimension 10 reads
\begin{eqnarray}\nonumber
\rho^0_{A}(s)&=&{17\over 2^{17}\cdot 5!\cdot 5!\cdot \pi^8}s^5
-{5\over 2^{22}\cdot 81\cdot \pi ^8}s^3\langle
g_s^2GG\rangle\\
 \nonumber &&+{1\over 2^{13}\cdot 9\cdot
\pi^6}s^3m_s({\frac{17}{48}\langle\bar
ss\rangle-\frac{7}{5}\langle\bar qq\rangle})\\
 \nonumber &&
+{1\over 2^{11}\cdot 9\cdot \pi^4}s^2( \frac {121}{20}{\langle\bar
qq\rangle}^2+7{\langle\bar qq\rangle}{\langle\bar
ss\rangle})\\
 \nonumber &&-{193\over 2^{18}\cdot 15\cdot \pi^6}s^2m_s\langle\bar
q\sigma\cdot Gq\rangle\\
 \nonumber &&
 +{1\over 2^{14}\cdot 27\cdot
\pi^4}s({1021\langle\bar qq\rangle+286\langle\bar
ss\rangle}){\langle\bar q\sigma\cdot Gq\rangle} \\
 &&+{5\over 2^6 \cdot
9\cdot \pi^2}m_s({-\langle\bar qq\rangle+\frac{13}{24}\langle\bar
ss\rangle}){\langle\bar qq\rangle}^2
\end{eqnarray}

We divide the zeroth moment from $\rho^0_{A}(s)$ by the D=6
corrections, which arise mainly from the four quark
condensate.
\begin{equation}\label{ww1} W^{I=0}_A(0,s_0)\sim
3.8\times 10^{-4} s^3_0+2\times 10^{-2} s_0 +1-{5.4\over
s_0}+{0.5\over s_0^2}\; .
\end{equation}
Here the D=8 term, which is the product of $\langle\bar qq
\rangle$ and $\langle g_s\bar q\sigma G q\rangle$, plays a very
important role. The convergence of the OPE requires $s_0\ge 13$
GeV$^2$, which renders the above FESR useless in the extraction of
$m_\Theta$. If we ignore the convergence criteria, we may arrive
at rather misleading results of $m_\Theta$ and $s_0$ as shown by
the variation of $m_\Theta$ with $s_0$ in Fig. 3 when power
corrections with D=6, 8, 10 are included.

\begin{figure}[hbt]
\begin{center}
\scalebox{0.8}{\includegraphics{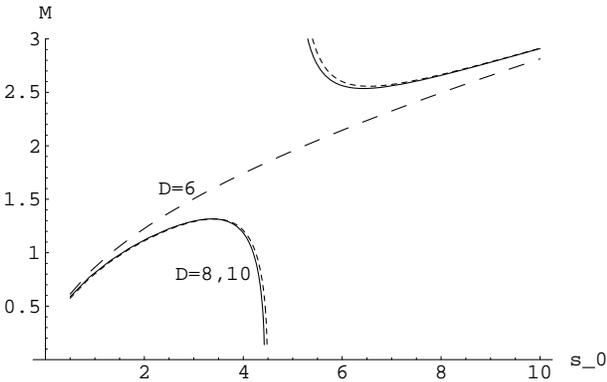}}
\end{center}\label{ai0m}
\caption{The variation of $M$ with $s_0$ corresponding to $\Pi_A$
of the I=0 current (\ref{current1}). The long-dashed, short-dashed
and solid curves correspond to $\rho^0_A$ with $D=6, 8, 10$
condensates respectively.  }
\end{figure}

\subsection{I=1 and I=2 Cases}

All the above analysis can be extended to $I=1, 2$ case. Roughly
speaking, the same conclusions hold. The spectral densities for
the I=1 case are
\begin{eqnarray}\nonumber
\rho^1_{B}(s) &=&{-1\over 2^{19}\cdot 175\cdot \pi^8}s^5 m_s\\
\nonumber && -{1\over 2^{16}\cdot 27\cdot \pi^6}s^4({\langle\bar
qq\rangle}-\frac{3}{4}{\langle\bar ss \rangle})\\
\nonumber &&-{1\over 2^{18}\cdot 15\cdot\pi^8}s^3m_s{\langle
g_s^2GG\rangle} \\
\nonumber && -{1\over 2^{12}\cdot 45\cdot \pi^6}s^3 \langle\bar
q\sigma\cdot Gq\rangle \\
\nonumber &&+{1\over 2^{12}\cdot 3 \cdot \pi^4}s^2m_s( \frac
{37}{6}{\langle\bar qq\rangle}-{\langle\bar
ss\rangle}){\langle\bar qq\rangle}\\ &&
 -{1\over
2^{8}\cdot 27\cdot \pi^2}s({24\langle\bar
qq\rangle}+29{\langle\bar ss \rangle}){\langle\bar qq\rangle}^2
\end{eqnarray}
and
\begin{eqnarray}\nonumber
\rho^1_{A}(s) &=&{11\over 2^{17}\cdot 5!\cdot 5!\cdot \pi^8}s^5
+{23\over 2^{22}\cdot 405 \cdot \pi ^8}s^3\langle
g_s^2GG\rangle\\
 \nonumber &&+{1\over 2^{17}\cdot
\pi^6}s^3m_s(\frac{11}{27}{\langle\bar ss\rangle}
-\frac{16}{15}{\langle\bar qq\rangle})
\\ \nonumber
&&+ {1\over 2^{11}\cdot 9 \cdot \pi^4}s^2( \frac
{27}{20}{\langle\bar qq\rangle}^2+3{\langle\bar
qq\rangle}{\langle\bar ss\rangle})\\
 \nonumber && -{89\over 2^{18}\cdot 15\cdot
\pi^6}s^2m_s\langle\bar
q\sigma\cdot Gq\rangle\\
 \nonumber &&
+{1\over 2^{13}\cdot 27\cdot \pi^4}s({135\langle\bar
qq\rangle+71\langle\bar ss\rangle}){\langle\bar q\sigma\cdot
Gq\rangle} \\
  &&+{1\over 2^6 \cdot 9\cdot \pi^2}m_s({-\langle\bar
qq\rangle+\frac{11}{24}\langle\bar ss\rangle}){\langle\bar
qq\rangle}^2 \; .
\end{eqnarray}

We also have normalized zeroth moments
\begin{equation}
W^{I=1}_B(0,s_0)\sim -4.4 \times 10^{-2} s_0+1-{25\over
s_0}+{66\over s_0^2}+{1406\over s_0^3}\; .
\end{equation}
\begin{equation}
W^{I=1}_A(0,s_0)\sim 7.7\times 10^{-4} s^3_0+7.3\times 10^{-3} s_0
+1-{5.1\over s_0}+{0.3\over s_0^2}\; .
\end{equation}

For I=2 case, the spectral densities read
\begin{eqnarray}\nonumber
\rho^2_{B} (s)& =&{-1\over 2^{11}\cdot 27\cdot
\pi^6}s^4{\langle\bar qq\rangle}\\
\nonumber &&-{1\over 2^{15}\cdot
9\cdot\pi^8}s^3m_s{\langle g_s^2GG\rangle}\\
\nonumber &&-{1\over 2^{10}\cdot 9\cdot \pi^6}s^3 \langle\bar
q\sigma\cdot Gq\rangle  \\
\nonumber && +{1\over 72 \cdot \pi^4}s^2m_s( {\langle\bar
qq\rangle}-\frac {3}{16}{\langle\bar ss\rangle}){\langle\bar
qq\rangle}\\
&& -{1\over 216\cdot \pi^2}s({24\langle\bar
qq\rangle}+23{\langle\bar ss \rangle}){\langle\bar qq\rangle}^2\;
,
\end{eqnarray}
\begin{eqnarray}\nonumber
\rho^2_{A} (s) &=&{7\over 2^{12}\cdot 5!\cdot 5!\cdot \pi^8}s^5
-{59\over 2^{18}\cdot 135\cdot \pi ^8}s^3\langle
g_s^2GG\rangle\\
 \nonumber &&+{1\over 2^{10}\cdot
3\pi^6}s^3m_s({\frac{7}{36}\langle\bar ss\rangle}-\frac
{4}{5}{\langle\bar qq\rangle})\\
\nonumber && + {1\over 2^6\cdot 9\pi^4}s^2( 2{\langle\bar
qq\rangle}^2+3{\langle\bar qq\rangle}{\langle\bar
ss\rangle})\\
 \nonumber &&-{499\over 2^{13}\cdot 45\cdot \pi^6}s^2m_s\langle\bar
q\sigma\cdot Gq\rangle\\
\nonumber && +{1\over 2^{8}\cdot 9\cdot \pi^4}s({751\over
12}\langle\bar qq\rangle+37{\langle\bar ss\rangle}){\langle\bar
q\sigma\cdot Gq\rangle}\\
 \nonumber &&+{1\over 18\cdot \pi^2}m_s({-\langle\bar
qq\rangle+\frac{11}{12}\langle\bar ss\rangle}){\langle\bar
qq\rangle}^2 \; .
\end{eqnarray}

Similarly we have
\begin{equation}
W^{I=2}_B(0,s_0)\sim 1-{6\over s_0}+{22\over s_0^2}+{505\over
s_0^3}\; .
\end{equation}
\begin{equation}
W^{I=2}_A(0,s_0)\sim 4.2\times 10^{-4} s^3_0+1.8\times 10^{-2} s_0
+1-{6.3\over s_0}+{0.1\over s_0^2}\; .
\end{equation}

For completeness we also present the variation of $m_\Theta$ with
$s_0$ in Figs. 4-7 although they are unreliable due to the lack of
convergence of OPE.

\begin{figure}[hbt]
\begin{center}
\scalebox{0.8}{\includegraphics{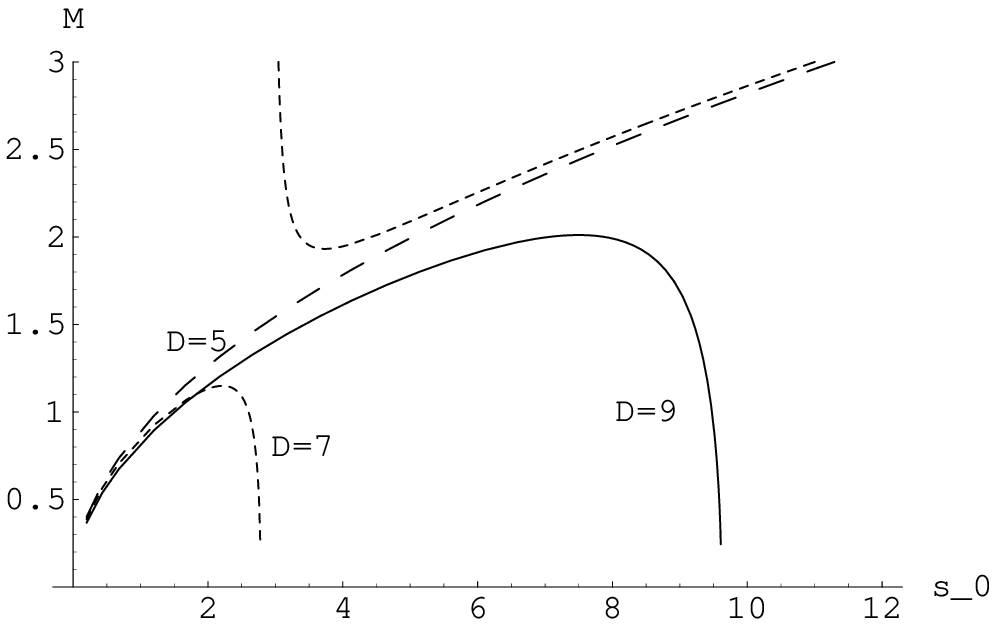}}
\end{center}\label{bi1m}
\caption{The variation of $M$ with $s_0$ corresponding to $\Pi_B$
of the I=1 current (\ref{current2}). }
\end{figure}

\begin{figure}[hbt]
\begin{center}
\scalebox{0.8}{\includegraphics{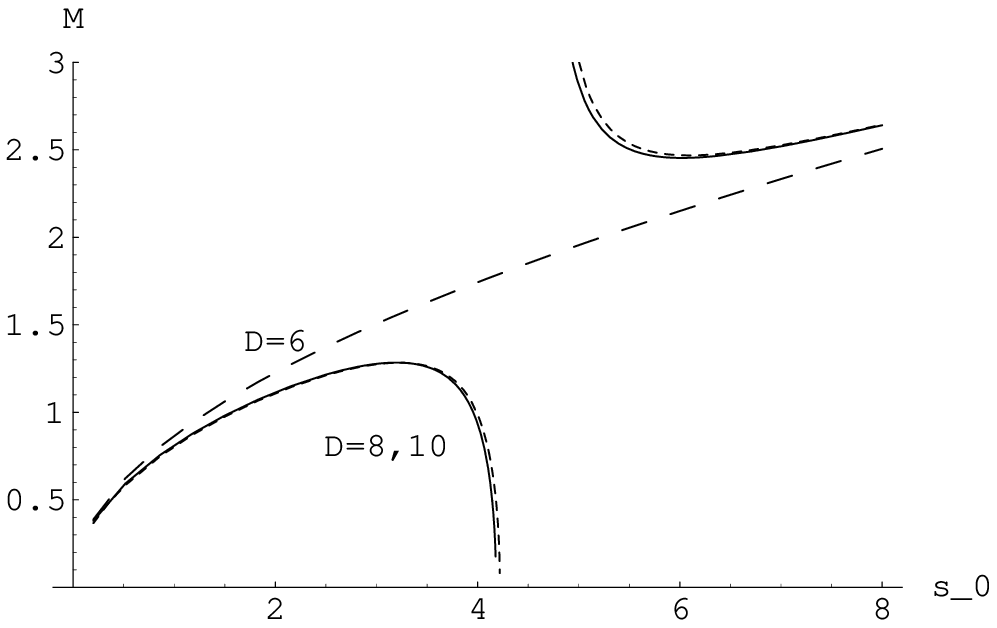}}
\end{center}\label{ai1m}
\caption{The variation of $M$ with $s_0$ corresponding to $\Pi_A$
of the I=1 current (\ref{current2}). }
\end{figure}

\begin{figure}[hbt]
\begin{center}
\scalebox{0.8}{\includegraphics{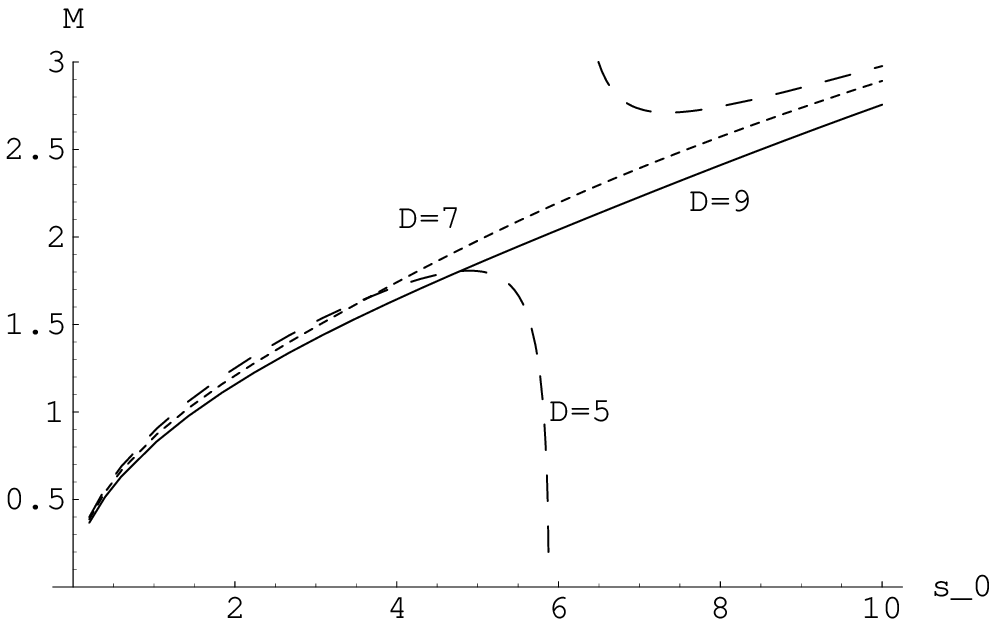}}
\end{center}\label{bi2m}
\caption{The variation of $M$ with $s_0$ corresponding to $\Pi_B$
of the I=2 current (\ref{current4}). }
\end{figure}

\begin{figure}[hbt]
\begin{center}
\scalebox{0.8}{\includegraphics{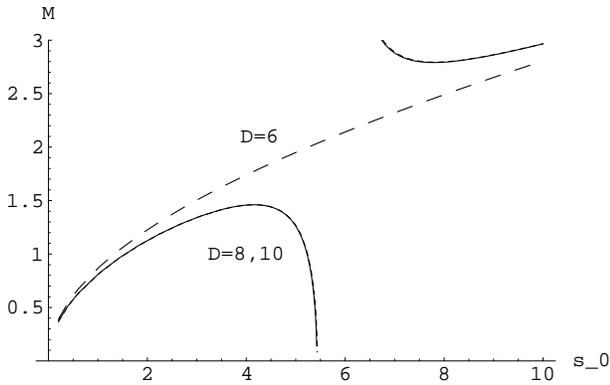}}
\end{center}\label{ai2m}
\caption{The variation of $M$ with $s_0$ corresponding to $\Pi_A$
of the I=2 current (\ref{current4}). }
\end{figure}

\section{Discussion}\label{sec4}

In this paper we have constructed the finite energy sum rules for
the spin 3/2 pentaquarks using the K-N color-octet type
interpolating currents. Both $g_{\mu\nu}$ and $g_{\mu\nu}{\hat p}$
tensor structures are unique for spin 3/2 pentaquarks. Because of
the high dimension of the pentaquark interpolating currents, power
corrections from condensates with $D\ge 5$ unfortunately turn out
to be numerically large for our three interpolating currents.
Especially the quark gluon mixed condensate plays a dominant role.

Numerical analysis indicates the stable region of the continuum
threshold is around several GeV$^2$. But OPE of both sum rules
converges only when the continuum threshold is very big, $s_0\ge
10$ GeV$^2$. Such a large value of $s_0$ is irrelevant to the
experimentally observed $\Theta^+$ pentaquark. In fact, lack of
OPE convergence renders these FESRs very sensitive to the high
dimension power corrections.

One may wonder whether the above conclusion is the artifact of
FESR approach only. We have also performed the numerical analysis
using the Laplace sum rule. Requiring the pole contribution is
greater than 40\% of the whole sum rule, we arrive at the lower
limit of the Borel parameter $M_{\mbox{min}}$. The convergence of
operator product expansion requires the high dimension operators
be suppressed. Numerically we may require the ratio between the
dimension D condensate and perturbative term is smaller than
$1/2^D$. In this way we get the upper limit of the Borel paramter
$M_{\mbox{max}}$. For all the above interpolating currents, we
find $M_{\mbox{min}}> M_{\mbox{max}}$ for both tensor structures.
In other words, there does not exist a working Borel window in the
Laplace sum rule analysis. This point has been noted for the spin
1/2 pentaquark case in \cite{narison}.

Recall both $\phi$ and $\omega$ mesons are narrow resonances above
threshold. $\phi$ decays into $K\bar K$ and $\omega$ decays into
$3\pi$. Their FESRs converge. The extracted vector meson masses
agree with the experimental data. The resonance pole contribution
dominates the $K\bar K$ or $3\pi$ continuum (background).
Similarly, if spin 3/2 pentaquarks {\sl really} exists as an
extremely narrow resonance as indicated by those positive
experiments, its pole contribution should dominate the $K N$
background. Hence one would expect (1) a converging FESR at least
for one of the two tensor structures; (2) a strong signal in the
working window with the continuum threshold slightly above
$m_\Theta^2$. But none of these spin 3/2 pentaquark FESRs
satisfies these conditions. This fact strongly indicates the
possible nonexistence of spin 3/2 pentaquarks around 1.53 GeV,
which is compatible with the most recent CLAS data \cite{clas}.
With a ten times larger database, the $nK^+$ spectrum is very
smooth around 1.53 GeV. In fact, CLAS found no signal of exotic
baryon resonances with $B=+1, S=+1$ up to 2.2 GeV.

Although our present investigation indicates the possible
nonexistence of a narrow spin 3/2 pentaquark using the
kaon-nucleon color-octet interpolating currents from QCD finite
energy sum rule analysis, this is not a {\sl strict} proof yet. An
exhaustive study of other interpolating currents in search of
excellent OPE convergence is very desirable. Only after OPE
convergence is established, may one be able to judge the existence
of pentaquarks rigorously from the behavior of QCD sum rules. One
important scheme is to include the coupled channel effects by
introducing a mixed interpolating current which is a combination
of several interpolating currents with different color structures.
The mixing and coupled channel effect may help suppress the high
dimension condensates and stabilize the sum rule. Work along this
direction is in progress.

\section*{Acknowledgments}

S.L.Z thanks A. Hosaka for helpful discussions. This project was
supported by the National Natural Science Foundation of China
under Grants 10375003 and 10421003, Ministry of Education of
China, FANEDD, Key Grant Project of Chinese Ministry of Education
(NO 305001) and SRF for ROCS, SEM.


\begin{thebibliography}{99}
\bibitem{lep} LEPS Collaboration, T. Nakano et al. , Phys. Rev. Lett. 91,
012002 (2003).
\bibitem{hicks}K. Hicks, hep-ex/0412048.
\bibitem{ddd}A. R. Dzierba, C. A. Meyer, A. P. Szczepaniak,
hep-ex/0412077.

\bibitem{slz2}Shi-Lin Zhu, Int. J. Mod. Phys. A 19, 3439(2004).
\bibitem{jw1} R. Jaffe and F. Wilczck, Phys. Rev.
Lett. 91, 232003 (2003).

\bibitem{dudek} J.J. Dudek and F.E. Close, Phys. Lett.B583, 278 (2004).
\bibitem{liww}W. W. Li et al., hep-ph/0312362, High Ener. Phys. Nucl.
Phys. 28, 918 (2004).
\bibitem{he}X.-G. He et al., Phys. Rev. D 71, 014006 (2005).

\bibitem{page}S. Capstick, P. R. Page and W. Roberts, Phys. Lett.
B 570, 185 (2003).
\bibitem{jaffe2}R. Jaffe and F. Wilczek, Phys. Rev. D 69, 114017
(2004).
\bibitem{faessler}T. Inoue et al., hep-ph/0407305.
\bibitem{takeuchi}S. Takeuchi and K. Shimizu, hep-ph/0410286.
\bibitem{kanada}Y. Kanada-En¡¯yo, O. Morimatsu and T. Nishikawa,
hep-ph/0404144.
\bibitem{zyzhang}F.Huang, Z.Y.Zhang and F.W.Yu, hep-ph/0411222.
\bibitem{hyodo}T. Hyodo and A. Hosaka, hep-ph/0502093, Phys. Rev.
D 71, 054017 (2005).
\bibitem{qsr-nishi}T. Nishikawa, Y. Kanada-En¡¯Yo, Y. Kondo and O.
Morimatsu, hep-ph/0411224.

\bibitem{qsr-zhu}Shi-Lin Zhu, Phys. Rev. Lett. 91, 232002 (2003).
\bibitem{qsr-mat1}R. D. Matheus etc, Phys. Lett. B 578 , 323 (2004).
\bibitem{qsr-oka}J. Sugiyama, T. Doi and M. Oka, Phys.
Lett. B 581,167(2004).
\bibitem{hpz}P. Z. Huang et al., Phys. Rev. D 69, 074004 (2004).
\bibitem{qsr-eide}M. Eide\"{u}ller, Phys. Lett. B597, 314 (2004).
\bibitem{qsr-mat2}R. D. Matheus, F. S. Navarra, M. Nielsen and R. R. da Silva,
 hep-ph/0406246, Phys. Lett. B 602, 185 (2004).
\bibitem{qsr-nava}F. S. Navarra et al., nucl-th/0408072, Phys. Lett. B 606, 335 (2005).
\bibitem{qsr-oga}B. L. Ioffe, A. G. Oganesian, hep-ph/0408152, JETP Lett. 80 (2004) 386.
\bibitem{qsr-mori1}T. Nishikawa et al., hep-ph/0410394,
Phys. Rev. D 71, 016001 (2005).
\bibitem{qsr-su}S. H. Lee, H. Kim, and Y. Kwon, hep-ph/0411104,
Phys. Lett. B609 (2005) 252.

\bibitem{narison}R. D. Matheus and S. Narison, hep-ph/0412063,
to appear in Nucl. Phys. B (Proc. Suppl.).
\bibitem{qsr-lee}H.-J. Lee, N. I. Kochelev and V. Vento,
hep-ph/0412127, Phys. Lett. B 610 (2005) 50.
\bibitem{wsl}Z.-G. Wang, W.-M. Yang, S.-L. Wan, hep-ph/0501015,
0501278, 0503007, 0503073.

\bibitem{svz}M. A. Shifman, A. I. Vainshtein and V. I. Zakharov, Nucl. Phys.
B 147,385 (1979).

\bibitem{fesr}N. V. Krasnikov et al., Z. Phys. C 26, 301 (1983);
S. Narison, {\sl QCD Sum Rules}, World Sci. Lect. Notes Phys. 26,
1 (1989).

\bibitem{ioffe} B. L. Ioffe, Nucl. Phys. B188, 317 (1981), Z. Phys. C 18 , 67 (1983).
\bibitem{reinders}L. J. Reinders, H.Rubinstein and S. Yazaki, Phys. Rept. 127 , 1 (1985).
\bibitem{ykc}K.-C. Yang et al., Phys. Rev. D 47, 3001 (1993).

\bibitem{clas}R. De Vita for the CLAS collaboration, talk at APS
meeting (April 16, 2005) at Tampa.
\end{thebibliography}
\end{document}